\documentclass[12pt,letterpaper]{article}

\usepackage[english]{babel}
\usepackage{amsmath}
\usepackage{amssymb}
\usepackage{amsfonts}
\usepackage{latexsym}
\usepackage[dvips]{graphicx}
\usepackage[dvips]{color}
\usepackage{rotating}
\usepackage{caption2}

\author{
\vspace{-0.7cm} \large S\'ebastien Roch\footnote{This work was done
while the first author was at \'Ecole Polytechnique, Montr\'eal.}\\
\vspace{-0.7cm} \normalsize Department of Statistics\\
\vspace{-0.7cm} \normalsize University of California, Berkeley\\
\vspace{-0cm} \normalsize Berkeley, CA
\and
\large \vspace{-0.7cm} Patrice Marcotte\\
\vspace{-0.7cm}\normalsize D\'epartement d'informatique\\
\vspace{-0.7cm}\normalsize et de recherche op\'erationnelle\\
\vspace{-0.7cm}\normalsize Universit\'e de Montr\'eal\\
\vspace{-0cm}\normalsize Montr\'eal, Qu\'ebec, Canada
\and
\vspace{-0.7cm} \large Gilles Savard\\
\vspace{-0.7cm} \normalsize D\'epartement de math\'ematiques\\
\vspace{-0.7cm} \normalsize et de g\'enie industriel\\
\vspace{-0.7cm} \normalsize \'Ecole Polytechnique de Montr\'eal\\
\vspace{-0cm} \normalsize Montr\'eal, Qu\'ebec, Canada
}
\title{An approximation algorithm \\
\vspace{-0.7cm}for Stackelberg network pricing}

\newcommand{\LL}{\mathcal{L}}
\newcommand{\TT}{\mathcal{T}}

\newcommand{\N}{\mathbf{N}}

\newcommand{\qed}{\begin{flushright} $\square$ \end{flushright}}

\newcommand{\UU}{\mathcal{U}}
\newcommand{\NN}{\mathcal{N}}
\newcommand{\SP}{\mathcal{SP}}
\newcommand{\LLL}{\mathcal{L}}
\newcommand{\init}{\mathrm{\textsc{init}}}
\newcommand{\term}{\mathrm{\textsc{term}}}

\newcommand{\bigkceil}{\Big\lceil \frac{k}{2}\Big\rceil}
\newcommand{\bigkfloor}{\Big\lfloor \frac{k}{2}\Big\rfloor}

\setcaptionwidth{13cm}

\begin{document}

\newtheorem{theorem}{Theorem}
\newtheorem{corollary}{Corollary}
\newtheorem{definition}{Definition}
\newtheorem{example}{Example}
\newtheorem{lemma}{Lemma}
\newtheorem{conjecture}{Conjecture}

\maketitle

\abstract{We consider the problem of maximizing the revenue raised from tolls set on the arcs of a transportation network, under the constraint that users are assigned to toll-compatible shortest paths. We first prove that this problem is strongly NP-hard. We then provide a polynomial time algorithm with a worst-case precision guarantee of $\frac{1}{2}\log_2 m_T+1$, where $m_T$ denotes the number of toll arcs. Finally we show that the approximation is tight with respect to a natural relaxation by constructing a family of instances for which the relaxation gap is reached.}

\medskip


\section{Introduction}

This paper focuses on a class of bilevel problems that arise
naturally when tariffs, tolls, or devious taxes are to be
determined over a network. This class of problems encompasses
several important optimization problems encountered in the
transportation, telecommunication, and airline industries. Our aim
is twofold: first, we show that the problem is NP-hard; then we
present a polynomial time algorithm with a tight worst-case
guarantee of performance.

Bilevel programming is a modelling framework for situations where
one player (the ``leader'') integrates within its optimization
schedule the reaction of a second player (the ``follower'') to its
own course of action. These problems are closely related to static
Stackelberg games and mathematical programs with equilibrium
constraints (or MPECs, see Luo, Pang and Ralph~\cite{LPR}), in
which the lower level solution characterizes the equilibrium state
of a physical or social system. Bilevel programs allow the
modelling of a variety of situations that occur in operations
research, economics, finance, etc. For instance, one may consider
the maximization of social welfare, taking into account the
selfish behavior of consumers. It is well-known that the taxation
of resources and services at marginal cost (Pigovian
taxes, see Pigou~\cite{Pigou}) maximizes global welfare. However, when some
resources fall outside the control of the leader, the social
optimum might not be reachable, yielding a ``second-best'' problem
of true Stackelberg nature (see Verhoef, Nijkamp and
Rietveld~\cite{Verhoef}, Hearn and Ramana~\cite{Hearn}, and
Larsson and Patriksson~\cite{Larsson} for traffic examples). In
constrast with these studies, we adopt the point of view of a firm
involved in the management of the network but oblivious to social
welfare; the firm's only goal is to maximize its own revenue.

\medskip

Bilevel programs are generally nonconvex and nondifferentiable,
i.e., to all practical extent, intractable. In particular, it has
been shown by Jeroslow~\cite{Jeroslow} that linear bilevel
programming is NP-hard. This result has been refined by Vicente,
Savard and J\'udice~\cite{Vicente}, who proved that obtaining a
mere certificate of local optimality is strongly NP-hard.
Actually, Audet et al.~\cite{Audet} unveiled the close
relationship between bilevel programming and integer programming.
This ``intractability'' has prompted the development of heuristics
that are adapted to the specific nature of the instance under
consideration, together with their worst-case analysis. Such
analysis was first performed for a network design problem with
user-optimized flows by Marcotte~\cite{MarcotteNDP}, who proved
worst-case bounds for convex optimization based heuristics. More
recently, worst-case analysis of Stackelberg problems has been
applied to job scheduling and to network design by
Roughgarden~\cite{Roughgarden,Roughgarden2}, to network routing by
Korilis, Lazar and Orda~\cite{Korilis} and to pricing of computer
networks by Cocchi et al.~\cite{Cocchi}. All these works focus on
``soft'' Stackelberg games, where the objectives of both players
are non-conflicting, and where heuristics are expected to perform
well in practice, although their worst-case behavior may turn out
to be bad.

In this paper, we analyze an approximation algorithm for the toll
optimization problem (\textsc{MaxToll} in the sequel) formulated
and analyzed by Labb\'e, Marcotte and Savard~\cite{LMS}. In this
game, which is almost zero-sum, a leader sets tolls on a subset of
arcs of a transportation network, while network users travel on
shortest paths with respect to the cost structure induced by the
tolls. Labb\'e et al.~\cite{LMS} proved that the Hamiltonian path
problem can be reduced to a version of \textsc{MaxToll} involving
{\sl negative} arc costs and {\sl positive} lower bounds on
tolls\footnote{It was also shown recently by Marcotte et
al.\cite{MarcotteSavardSemet} that the TSP is a special case of
{\sc MaxToll}.}. In this paper, we improve this result by showing
that \textsc{MaxToll}, without lower bound constraints on tolls,
is strongly NP-hard. Next, in the single-commodity case, we
provide a polynomial time algorithm with a performance guarantee
of $\frac{1}{2}\log_2 m_T + 1$, where $m_T$ denotes the number of
toll arcs in the network. We then use this result as well as
specially constructed instances to prove the tightness of our
analysis, as well as the optimality of the approximation factor
obtained with respect to a natural upper bound.

The rest of the paper is organized as follows. In
Section~\ref{section:preliminaries}, we state the problem and
prove that it is NP-hard. In Section~\ref{section:algorithm} we
introduce an approximation algorithm whose performance is analyzed
in Section~\ref{section:analysis}.

\section{The model and its complexity} \label{section:preliminaries}

\subsection{The model}
The generic \emph{bilevel toll problem} can be expressed as

$$\max_{\mathbf{T}} \mathbf{Tx}$$
where $\mathbf{x}$ 
is the partial solution of the parametric linear program
\begin{eqnarray*}
\min_{\mathbf{x},\mathbf{y}} && (\mathbf{c}_1+\mathbf{T})\mathbf{x} 
+ \mathbf{c}_2\mathbf{y}\\
\hbox{s.t.} && A_1\mathbf{x}+A_2 \mathbf{y}=\mathbf{b}\\
&& \mathbf{x},\mathbf{y}\ge 0,
\end{eqnarray*}
In the above, $\mathbf{T}$ represents a \emph{toll vector}, $\mathbf{x}$ the vector
of \emph{toll commodities} and $\mathbf{y}$ the vector of \emph{toll-free
commodities}.

We shall consider a combinatorial version of this problem. Let
$G=(V,A)$ be a directed multigraph with two distinguished
vertices: the origin $s\in V$ and the destination $t\in V$. The
arc set $A$ is partitioned into subsets $A_T$ and $A_U$ of
\emph{toll} and \emph{toll-free} arcs, of respective cardinalities
$m_T$ and $m_U$. Arcs are assigned \emph{fixed costs} $c:A_T \to
\N^{m_T}$ and $d:A_U \to \N^{m_U}$ (in the sequel, $\N=\{0,\, 1,\, \ldots\}$).
Once \emph{tolls} are added to the fixed costs of $A_T$, we obtain
a \emph{toll network} $\NN_T=(G,\,c+T,\, d,\, s,\, t)$. Denoting
by $\SP[\NN_T]$ the set of shortest paths from $s$ to $t$, we can
then formulate \textsc{MaxToll} as the combinatorial mathematical
program (see also Figure~\ref{figure:gareyjohnson}):
\begin{equation}
\max_{T \ge 0\atop P\in \SP[\NN_T]}
\sum_{e\in A_T\cap P}T(e).
\end{equation}
This is a single-commodity instance of
 the toll setting problem analyzed by Labb\'e et al.~\cite{LMS}.

\begin{figure}[h]
\begin{center}
\framebox{
\begin{minipage}{14cm}
\begin{tabular}{ll}
\textbf{INSTANCE}: & -- a directed graph $G=(V,\,A_T\cup A_U)$\\
& -- fixed cost vectors  $c:A_T\to \N^{m_T}$ and \mbox{$d:A_U \to \N^{m_U}$}\\
& -- distinguished vertices $s,t \in V$ such that there \\
& \phantom{--} exists a simple path from $s$ to $t$ in $(V,\, A_U)$\\
\\
\textbf{SOLUTION}: & -- a nonnegative toll vector $T$ on $A_T$\\
& -- a simple $s-t$ path $P$ of minimal length  w.r.t.\\
& \phantom{--} the cost structure $(c+T,\, d)$\\
\\
\textbf{MEASURE}: & -- maximize $\sum_{e\in A_T\cap P}T(e)$
\end{tabular}
\end{minipage}
} \caption{\textsc{MaxToll}} \label{figure:gareyjohnson}
\end{center}
\end{figure}

\begin{figure}
\begin{center}
\input{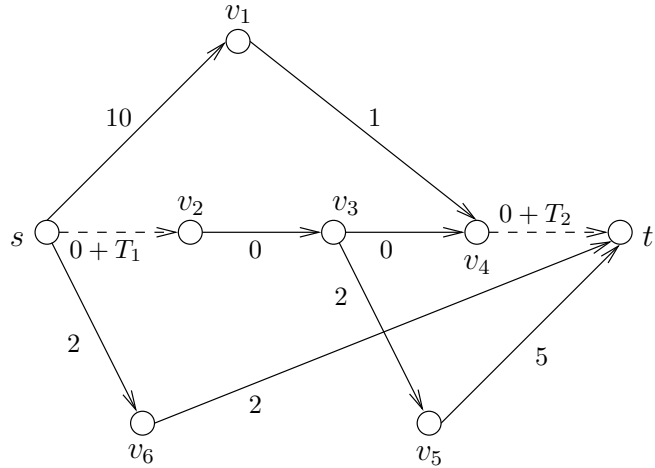}
\end{center}
\caption{This network contains two toll arcs (represented by
dashed arcs). Fixed costs are given by numbers close to each arc.
The optimal path is $(s,\, v_2,\, v_3,\, v_4,\, t)$ with a revenue of 4
when $T_1=2$ and $T_2=2$.}
  \label{figure:exemple}
\end{figure}

In this framework, the leader must strike the right balance
between low toll levels, which generate low revenue, and high
levels, which could also result in low revenue, as the follower
would select a path with few toll arcs, or even none. An instance
of \textsc{MaxToll} is illustrated in Figure~\ref{figure:exemple}.

Several remarks are in order. Firstly, to avoid a trivial situation,
we posit the existence of at least one toll-free path from $s$ to
$t$.  Secondly, our formulation implies that, given ties at the
lower level (the user's level), the leader chooses among the
toll-compatible shortest paths the one travelled by the follower.
Note that a risk-averse leader could always force the use of the
most profitable path by subtracting a small amount from every toll
on that path, thus yielding a revenue as close as desired to the
revenue generated by the ``cooperative'' solution. Thirdly, once a
path $P$ has been selected by the leader, toll arcs outside $P$
become irrelevant. In practice, the removal of these arcs can be
achieved by setting tolls to an arbitrarily large value on toll
arcs outside $P$. We denote by $\NN_T(P)$ the network where these
arcs have been removed. Finally, our central results hold also in
a version of {\sc MaxToll} where $T$ is unconstrained; in this
case, negative tolls can be interpreted as subsidies. Actually,
Labb\'e, Marcotte and Savard~\cite{LMS2} have constructed
instances where the optimal solution involves negative tolls.
Nevertheless, throughout most of the paper we focus on nonnegative
tolls because (i) this case is interesting in its own sake, (ii)
intermediate results are easier to interpret when tolls are
thought to be nonnegative.

\medskip

A natural upper bound on the leader's revenue has been derived
by Labb\'e et al. \cite{LMS} using duality arguments from linear programming
theory. It also follows from Theorem~\ref{theorem:characterization}
of Section~\ref{section:characterizing}.
Let $\LLL_{\infty}$ be the length of  a shortest toll-free
path and let $\LLL(P)$ be the length of a given path $P$ with 
$T(e) \equiv 0$ for all toll arcs $e$.
\begin{theorem}
Let $P$ be a path. Then the optimal revenue associated with $P$ is
bounded above by
\begin{equation}\label{UpperBound}
B(P)\equiv\LLL_{\infty} - \LLL(P).
\end{equation}
\end{theorem}
Since $\LLL_{\infty}$ does not depend on $P$, it follows that the
largest upper bound corresponds to the path with smallest value of
$\LLL(P)$; that is, $P$ is a shortest path when tolls are set to
0. We denote the length of such a path by $\LLL_{0}$ and by
\begin{equation}
LP=\LLL_{\infty}-\LLL_{0}
\end{equation}
the value of a path-independent upper bound. This bound is simply
the difference between the costs of shortest paths corresponding
to infinite and null tolls, respectively. Note that, if the set of
toll arcs is a singleton, the upper bound can always be achieved.

\subsection{NP-hardness of {\sc MaxToll}}

The purpose of this section is to show that \textsc{MaxToll} is
strongly NP-hard. We also prove that a version of \textsc{MaxToll}
where the toll vector is unconstrained shares this property, thus settling
a conjecture of Labb\'e et al.~\cite{LMS}
about the complexity status of the generic toll setting
problem.

\begin{theorem}\label{thm:nphard}
\textsc{MaxToll} is strongly NP-hard.
\end{theorem}
\emph{Proof:} Let $C$ denote the sum of all fixed costs. It is not
difficult to show that there exists an optimal toll vector $T$
that is integer-valued and less than $C+1$; in particular, optimal
solutions are of polynomial size.

Now, consider a reduction from 3-SAT to \textsc{MaxToll} (see
\cite{GJ}). Let $x_1,\ldots,\, x_n$ be $n$ Boolean variables
and
\begin{equation}
F=\bigwedge_{i=1}^{m} (l_{i1}\lor l_{i2}\lor l_{i3})
\end{equation}
be a 3-CNF formula consisting of $m$ clauses with literals (variables or
their negations) $l_{ij}$. For each clause, we construct
a sub-network comprising one toll arc
for each literal as shown in Figure~\ref{figure:cell}.

\begin{figure}
\begin{center}
\input{cellule.pstex_t}
\end{center}
\caption{Sub-network for clause $(l_{i1}\lor l_{i2}\lor l_{i3})$.}
  \label{figure:cell}
\end{figure}

The idea is the following: if the optimal path goes through toll
arc $T_{ij}$, then the corresponding literal $l_{ij}$ is \textsc{true} (note:
if $l_{ij} = \overline{x}_k$, then $x_k=$\textsc{false}). The sub-networks
are connected by two arcs, a toll-free arc of cost 2 and
a toll arc of cost 0, as shown in Figure~\ref{figure:np}.

\setcaptionwidth{15cm}
\begin{figure}
\begin{center}
\begin{sideways}
\input{np.pstex_t}
\end{sideways}
\end{center}
\caption{Network for the formula $(x_1\lor x_2\lor\overline{x}_3)
\land(\overline{x}_2\lor x_3\lor\overline{x}_4)
\land(\overline{x}_1\lor x_3\lor x_4)$. Inter-clause arcs
are bold. Path through $T_{12},\, T_{22},\, T_{32}$
is optimal ($x_2=x_3=$\textsc{true}).}
\label{figure:np}
\end{figure}
\setcaptionwidth{13cm}

If $F$ is satisfiable, we want the optimal path to go through a
single toll arc per sub-network (i.e., one \textsc{true} literal
per clause) and simultaneously want to make sure that the
corresponding assignment of variables is consistent; i.e., paths
that include a variable and its negation must be ruled out. For
that purpose, we assign to every pair of literals corresponding to
a variable and its negation an inter-clause toll-free arc between
the corresponding toll arcs (see Figure~\ref{figure:np}). As we
will see, this implies that \emph{inconsistent} paths, involving a
variable and its negation, are suboptimal.

\medskip

 Since the  length of a shortest toll-free path is $m+2(m-1)=3m-2$ and
that of a shortest path with zero tolls is 0,  $3m-2$ is an upper
bound on the revenue. We claim that $F$ is satisfiable if and only
if the optimal revenue is equal to that bound.

\medskip

Assume that the optimal revenue is equal to $3m-2$. Obviously, the
length of the optimal path when tolls are set to 0 must be 0,
otherwise the upper bound cannot be reached. To achieve this, the
optimal path has to go through one toll arc per sub-network (it
cannot use inter-clause arcs) and tolls have to be set to 1 on
selected literals, $C+1$ on other literals and 2 on tolls $T_k,\
\forall\ k$. We claim that the optimal path does not include a
variable and its negation. Indeed, if that were the case, the
inter-clause arc joining the corresponding toll arcs would impose
a constraint on the tolls between its endpoints. In particular,
the toll $T_k$ immediately following the initial vertex of this
inter-clause arc would have to be set at most to 1, instead of 2.
This yields a contradiction. Therefore, the optimal path must
correspond to a consistent assignment, and $F$ is satisfiable
(note: if a variable and its negation do not appear on the optimal
path, this variable can be set to any value).

\medskip

Conversely if $F$ is satisfiable, at least one literal per clause
is \textsc{true} in a satisfying assignment.
Consider the path going through the toll arcs
corresponding to these literals. Since the assignment is
consistent, the path does not simultaneously include a variable and its negation,
and no inter-clause arc limits the revenue. Thus,
the upper bound of $3m-2$ is reached on this path.

\medskip

Finally, note that the number of arcs in the reduction is less than
\begin{equation*}
10m + 2(m-1) + (3m)^2
\end{equation*}
 and that all constants are polynomially bounded in $m$.\qed

It is not difficult to prove that the same NP-hardness
reduction works when negative tolls are allowed.
\begin{theorem}
\textsc{MaxToll} is still strongly NP-hard when negative tolls are allowed.
\end{theorem}
\emph{Proof:} We use the same reduction as in the nonnegative
case. The proof rests on two results proved in~\cite{LMS}. First,
the upper bound is  valid when $T$ is unrestricted. Second, there
exists optimal solutions of polynomial size. The latter result
follows from a polyhedral characterization of the feasible set.

\medskip

From the first result, we know that the $3m-2$ upper bound on the revenue
is unchanged. On the other hand, if $F$ is satisfiable, the feasible solution considered
in the nonnegative case still yields a $3m-2$ revenue. We only have to make
sure that negative tolls cannot produce a $3m-2$ revenue when $F$ is
not satisfiable. Again, to reach the upper bound, one has to use
a path of length 0 when tolls are set to 0. Consequently the optimal path
comprises exactly one literal per clause. Now the toll-free arcs of length 1 and 2
limit the $T_{ij}$'s on the path to 1 and the $T_k$'s to 2, so negative
tolls are useless in this case. Indeed, inconsistent paths will see their
revenue limited by inter-clause arcs without any possibility to make up for
the loss incurred from negative tolls.\qed

\section{Approximation algorithm} \label{section:algorithm}

In this section, we devise a polynomial-time approximation
algorithm for \textsc{MaxToll}. Such an algorithm is guaranteed to
compute a feasible solution with objective at least $OPT/\alpha$,
where $OPT$ is the optimal revenue and $\alpha$, which depends on
the number $m_T$ of toll arcs, denotes the approximation factor.
For a survey of recent results on approximation algorithms, the
reader is referred to Hochbaum~\cite{Hochbaum},
Vazirani~\cite{Vazirani} and Ausiello et al.~\cite{Ausiello}.

\subsection{Preliminaries: characterizing consistent tolls}\label{section:characterizing}

The leader is only interested in paths that have the potential of
generating positive revenue. This remark warrants the following
definitions. Recall that $\NN_T(P)$ is the network $\NN_T$ in
which toll arcs outside the path $P$ have been removed.
\begin{definition}
Let $m_T^P$ denote the number of toll arcs in a path $P$ from $s$
to $t$ . We say that $P$ is {\bf valid} if $m_T^P\geq 1$ and $P$
is a shortest path with respect to a null toll vector $T$, i.e.,
$P\in \SP[\NN_0(P)]$.
\end{definition}
It is clear  that non-valid paths cannot generate revenue. Any
valid path $P$ can be expressed as a sequence
\begin{equation}
P=\big(\upsilon_{0,1},\, \tau_1,\, \upsilon_{1,2},\, \tau_2,
\ldots ,  \tau_{m_T^P},\, \upsilon_{m_T^P,m_T^P+1}\big)
\end{equation}
where $\tau_i$ is the $i$-th toll arc of $P$ (in the order of
traversal) and $\upsilon_{i,i+1}$ is the toll-free subpath of $P$
from the terminal vertex $\term(\tau_i)$ of $\tau_{i}$ to the
initial vertex $\init(\tau_{i+1})$ of $\tau_{i+1}$. According to
this notation, $\upsilon_{0,1}$ starts in $s$ and
$\upsilon_{m_T^P,m_T^P+1}$ ends in $t$. Since $P\in
\SP[\NN_0(P)]$, $\upsilon_{i,i+1}$ is a shortest toll-free path
from $\term(\tau_{i})$ to $\init(\tau_{i+1})$. We extend this
notation to $\upsilon_{i,j}$, a shortest toll-free path from
$\term(\tau_{i})$ to $\init(\tau_{j})$, with length $\UU_{i,j}$.
For $k<l$, let $\LLL_{k,l}$ be the length of $P$ (i.e. the sum of costs) 
from $\term(\tau_{k})$ to $\init(\tau_{l})$ with tolls set to 0, and
$\TT_{k,l}$ be the sum of tolls between $\term(\tau_{k})$ and
$\init(\tau_{l})$ on $P$,
\begin{equation}
\LLL_{k,l} = \sum_{i=k}^{l-1}\UU_{i,i+1} +
\sum_{i=k+1}^{l-1}c(\tau_{i}) \qquad \TT_{k,l} =
\sum_{i=k+1}^{l-1}T(\tau_{i}),
\end{equation}
with the convention that $\sum_{i=k}^l x_i=0$ if $l<k$.

\begin{figure}
\begin{center}
\input{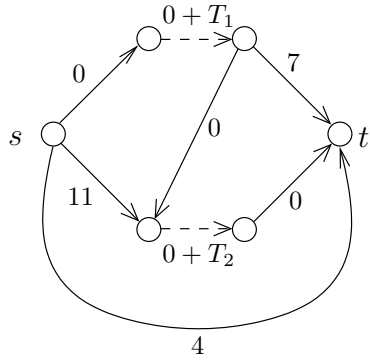}
\end{center}
\caption{Equivalent representation of the example of Figure~\ref{figure:exemple}.
All arcs have been replaced by shortest paths between toll arcs and
origin/destination.}
  \label{figure:canonique}
\end{figure}

\begin{definition}
Let $P$ be a valid path. The toll vector $T$ is {\bf consistent}
with $P$ if $P\in \SP[\NN_T(P)]$; that is, $P$ remains a
shortest path when tolls outside $P$ are removed and tolls on $P$
are set according to the vector $T$.
\end{definition}
The following result, which characterizes consistent tolls, is the
starting point of our algorithm.
\begin{theorem}\label{theorem:characterization}
Let $P$ be a valid path. Then, the toll vector $T$ is consistent
with $P$ if and only if
\begin{equation}\label{eq:characterization}
\LL_{i,j} + \TT_{i,j} \leq \UU_{i,j}\qquad\forall\ 0\leq i < j\leq
m_T^P+1.
\end{equation}
\end{theorem}
Before providing the proof of this theorem, we give an example of
its application. Consider again the network of Figure~\ref{figure:exemple}.
Take the path $P = (s,\,v_2,\,v_3,\,v_4,\,t)$. Using the renumbering system
introduced above, we have that $\tau_1 = (s,\,v_2)$,
$\tau_2 = (v_4,\,t)$, $v_{0,1}$ is empty (it actually 
consists in the node $s$ by itself), 
$v_{1,2} = (v_2,\,v_3,\,v_4)$ and $v_{2,3}$ is also
empty. Other toll-free subpaths that play a role
in Theorem~\ref{theorem:characterization} are
$v_{0,2} = (s,\,v_1,\,v_4)$, 
$v_{1,3} = (v_2,\,v_3,\,v_5,\,t)$, and
$v_{0,3} = (s,\,v_6,\,t)$. The corresponding lengths are 
$\UU_{0,1} = \UU_{1,2} = \UU_{2,3}= 0$, 
$\UU_{0,2} = 11$, $\UU_{1,3} = 7$, and
$\UU_{0,3} = 4$. Thus, the inequalities
in Theorem~\ref{theorem:characterization} are
\begin{equation*}
T(\tau_1) \leq 11,\qquad
T(\tau_2) \leq 7,\qquad
T(\tau_1) + T(\tau_2) \leq 4. 
\end{equation*}
From this we conclude that one possible set of optimal
taxes is $T(\tau_1) = T(\tau_2) = 2$ as claimed in 
Figure~\ref{figure:exemple}.
Figure~\ref{figure:canonique} provides an equivalent representation
of this example according to
Theorem~\ref{theorem:characterization}.

\begin{figure}
\begin{center}
\input{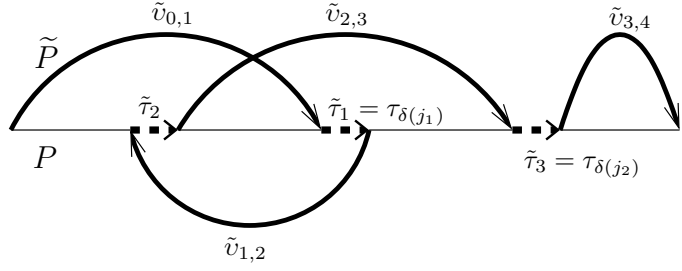}
\end{center}
\caption{The thick path is $\tilde P$ while the straight
thin path is $P$. An example of the ``backward'' toll-free
subpaths referred to in the proof is
$\tilde{v}_{1,2} \to \tilde{\tau}_2 \to \tilde{v}_{2,3}$. This gets
replaced by the section of $P$ between $\tau_{\delta(j_1)}$
and $\tau_{\delta(j_2)}$, which is clearly shorter.}
\label{figure:theorem4}
\end{figure}

\emph{Proof of Theorem~\ref{theorem:characterization}:} 
$\Longrightarrow )$ Obvious from the very
definition of consistent tolls. $\Longleftarrow )$ 
We need to check that the length of $P$ 
remains smaller than or equal to
the length of any other path when tolls satisfy
(\ref{eq:characterization}).
This
looks rather obvious. However, one must be careful about paths
that borrow a subset of the toll arcs of $P$, possibly 
in a \emph{different} sequence (note that toll-free paths
are taken care of by setting $i=0$ and $j=m_T^P + 1$ above).
Assume, by contradiction, that such a path $\widetilde{P}$ is
strictly shorter than $P$ in $\NN_T(P)$, and that conditions
(\ref{eq:characterization}) are satisfied. We note
\begin{equation}
\widetilde{P}=\big(\tilde{\upsilon}_{0,1},\, \tilde{\tau}_1,\,
\tilde{\upsilon}_{1,2},\, \ldots ,
\tilde{\tau}_{m_T^{\widetilde{P}}},\,
\tilde{\upsilon}_{m_T^{\widetilde{P}},m_T^{\widetilde{P}}+1}\big)
\end{equation}
with corresponding $\widetilde{\UU}_{i,j}$,
$\widetilde{\LLL}_{k,l}$  and $\widetilde{\TT}_{k,l}$ for all $i,
\,j,\, k,\, l$ with $k<l$. For all $i$,
$\tilde{\tau}_i=\tau_{\delta(i)}$ for some injective function
$\delta$ (toll arcs outside $P$ are irrelevant). To
derive a contradiction, we will construct a path that is shorter
than $\widetilde{P}$ by modifying the ``backward'' toll-free
subpaths of $\widetilde{P}$. This improved path will happen to be $P$.
See Figure~\ref{figure:theorem4} for an illustration.

Let  $j_1=1$ and $ j_k = \min\{j>j_{k-1}:
\delta(j)>\delta(j_{k-1})\}$, as long as such $j_k$ exists, the last
one being denoted $j_K$. We further define $\delta(0)=j_0=0$ and
$\delta(m_T^{\widetilde{P}}+1)=j_{K+1}=m_T^{\widetilde{P}}+1$.
We claim that
\begin{equation*}
\LLL_{\delta(j_{k-1}),\delta(j_k)} +
\TT_{\delta(j_{k-1}),\delta(j_k)}
\leq \widetilde{\LLL}_{j_{k-1},j_k} + \widetilde{\TT}_{j_{k-1},j_k},
\end{equation*}
for all $k$. This implies that
we can replace the subpath of $\widetilde{P}$ between
$\term(\tilde{\tau}_{j_{k-1}})$ and $\init(\tilde{\tau}_{j_k})$,
by the subpath of $P$ between $\term(\tau_{\delta(j_{k-1})})$  and
$\init(\tau_{\delta(j_k)})$ without increasing its length. But
after doing this for all $k$, we obtain $P$ which is a
contradiction. We divide the claim above in two cases.
Firstly, if $j_k-1 =j_{k-1}$ (i.e. $\widetilde{P}$ ``goes in the
direction of $P$'', e.g. $\tilde{v}_{0,1}$ in 
Figure~\ref{figure:theorem4}), then
\begin{eqnarray*}
\widetilde{\LLL}_{j_{k-1},j_k} + \widetilde{\TT}_{j_{k-1},j_k}
= \UU_{\delta(j_{k-1}),\delta(j_k)} \geq \LLL_{\delta(j_{k-1}),\delta(j_k)} +
\TT_{\delta(j_{k-1}),\delta(j_k)},
\end{eqnarray*}
where we have used (\ref{eq:characterization})
(note that in this case, there are no toll arcs on $\widetilde{P}$ between
$\term(\tilde{\tau}_{j_{k-1}})$ and $\init(\tilde{\tau}_{j_k})$).
Otherwise, $j_k-1 \neq j_{k-1}$ (i.e. $\widetilde{P}$ ``goes in the
direction opposite to $P$'', e.g. 
$\tilde{v}_{1,2}\to \tilde{\tau}_2 \to \tilde{v}_{2,3}$ in 
Figure~\ref{figure:theorem4}).
By definition of $j_k$, we have $\delta(j_k-1)\leq
\delta(j_{k-1})$ for all $1\leq k\leq K+1$. For example,
in Figure~\ref{figure:theorem4}, we have
$\tau_{\delta(j_2-1)} = \tilde{\tau}_2$ which clearly comes
before $\tau_{\delta(j_2)} = \tilde{\tau}_3$ on $P$.
From this and (\ref{eq:characterization}), we get
\begin{eqnarray*}
\widetilde{\LLL}_{j_{k-1},j_k} + \widetilde{\TT}_{j_{k-1},j_k}
&=& \widetilde{\LLL}_{j_{k-1},j_k-1} + \widetilde{\TT}_{j_{k-1},j_k-1}
+ c(\tau_{\delta(j_k-1)}) + T(\tau_{\delta(j_k-1)}) + \UU_{\delta(j_k-1),\delta(j_k)}\\
&\geq& \UU_{\delta(j_k-1),\delta(j_k)} \geq
\LLL_{\delta(j_k-1),\delta(j_k)} + \TT_{\delta(j_k-1),\delta(j_k)}\\
&\geq& \LLL_{\delta(j_{k-1}),\delta(j_k)} +
\TT_{\delta(j_{k-1}),\delta(j_k)}.
\end{eqnarray*}
\qed

\medskip

{\bf Remark.} There always exists an optimal solution with tolls
less than $C+1$ (see Theorem~\ref{thm:nphard} for a definition
of $C$). Indeed, $\UU_{i,j}<C+1,\ \forall\, i,j$ implies
that  optimal tolls on $P$ have to be lower than $C+1$. Therefore,
fixing tolls to $C+1$ outside $P$ generates no additional active
constraints on the tolls of $P$.

\subsection{High-level description of the algorithm}

\textsc{ExploreDescendants} is an approximation algorithm
motivated by the previous characterization of consistent tolls. Initially,
the algorithm computes the optimal revenue associated with a
shortest path in $\SP[\NN_0]$, i.e, a path with the largest upper
bound $B(P)$\footnote{As shown in the next section, this can be
achieved in polynomial time.}. If revenue is smaller than the
upper bound $LP$, Theorem~\ref{theorem:characterization} implies
that there exists a toll-free subpath $\upsilon_{k,l}$ $k<l$ whose
short length forces some tolls in $P$ to be small. To relax this
constraint, it makes sense to skip the subpath of $P$ between
$\term(\tau_k)$ and $\init(\tau_l)$ and to replace it by
$\upsilon_{k,l}$. This yields a new path whose length will not be
much larger than the length of $P$, and for which some
constraints~(\ref{eq:characterization}) have been removed. This
path is a natural candidate for improved revenue. By repeating
this process, we will show that
Algorithm~\textsc{ExploreDescendants} can uncover, in polynomial
time, a path with good approximation properties.

\medskip

Algorithm~\textsc{ExploreDescendants} is streamlined in
Figure~\ref{figure:exploredescendants}. It comprises two
subroutines, \textsc{MaxRev} and \textsc{TollPartition}.
\textsc{MaxRev} computes the largest revenue compatible with the
shortest path status of $P$.
 Starting from $P$, \textsc{TollPartition} generates two descendants
 of path $P$.
The algorithm is initialized with $P_0$ in $\NN_0$, a shortest
path of length $\LLL_0$.
\begin{figure}[h]
\begin{center}
\framebox{
\begin{minipage}{15cm}
\textbf{Algorithm} \textsc{ExploreDescendants}

\textbf{Input:} a path $P$

\textbf{Output:} a path $\overline{P}$, tolls $\overline{T}$ and
objective value $\overline{V}$

\begin{itemize}
\item Compute maximum revenue $V_P$ achievable on $P$ and
corresponding toll vector $T_P$: $\left(V_P,T_P,\{(i'(k),j'(k))\}_{k=1}^{m_T^P}\right):=$
\textsc{MaxRev}$(P)$
\item If $V_P<B(P)$ then
\begin{itemize}
\item Derive new paths from $P$ :
\begin{equation*}
(P_1,\, P_2)
:=  \mathrm{\textsc{TollPartition}}
\left(P, \{(i'(k),j'(k))\}_{k=1}^{m_T^P}\right)
\end{equation*}
\item For $i=1,2$ :  $(\overline{V}_i,\, \overline{T}_i,\,
\overline{P}_i):=$ \textsc{ExploreDescendants}$(P_i)$
\item Return $\overline{V}:=\max\{V_P,\,\overline{V}_1,\,  \overline{V}_2\}$
and corresponding toll vector $\overline{T}$ and path
$\overline{P}$
\end{itemize}
\item Else return $(\overline{V},\, \overline{T},\, \overline{P}):=(V_P,\, T_P,\, P)$
\end{itemize}
\end{minipage}
}
\caption{Algorithm \textsc{ExploreDescendants}} \label{figure:exploredescendants}
\end{center}
\end{figure}

\clearpage

\subsection{Maximizing path-compatible revenue}

Let $P$ be a valid path in $\NN$ denoted
by
\begin{equation}
P=\big(\upsilon_{0,1},\, \tau_1,\, \upsilon_{1,2},\, \ldots ,
\tau_{m_T^P},\,  \upsilon_{m_T^P,m_T^P+1}\big)
\end{equation}
with corresponding values of $\UU_{i,j}$, $\LLL_{k,l}$ and
$\TT_{k,l}$ for all $i,\, j,\, k,\, l$, with $k<l$. The optimal tolls
compatible with $P$'s shortest path status can be obtained from a simple
greedy algorithm (see Figure~\ref{figure:maxrev}): consider each toll
arc in order of traversal, and fix its value to the largest value
allowed by Theorem~\ref{theorem:characterization}, taking into
account the tolls previously set. 
This leads to the
recursion
\begin{equation}\label{eq:optimaltaxes}
T(\tau_k):=t_k\equiv \min_{0\leq i<k<j\leq m_T^P+1}
\left\{\UU_{i,j}-\LLL_{i,j}-\sum_{l=i+1}^{k-1} t_l \right\}.
\end{equation}
The
validity of the above formula rests on Theorem~\ref{lemma:main}
where we construct a sequence of toll-free subpaths such that (i)
every toll of $P$ is bounded from above by at least one path in
the sequence (condition (\ref{eq:covering}) below),
(ii) the sum of the tolls defined
by~(\ref{eq:optimaltaxes}) is equal to the sum of the bounds
imposed by the paths of the sequence (condition (\ref{eq:main}) below).
\begin{figure}[h!]
\begin{center}
\framebox{
\begin{minipage}{15cm}
\textbf{Algorithm} \textsc{MaxRev}

\textbf{Input:} a path $P$

\textbf{Output:} tolls $T_P$,
objective value $V_P$, and 
sequence of indices $\{(i'(k),j'(k))\}_{k=1}^{m_T^P}$

\begin{itemize}
\item Denote
$P=\big(\upsilon_{0,1},\, \tau_1,\, \upsilon_{1,2},\, \tau_2,
\ldots ,  \tau_{m_T^P},\, \upsilon_{m_T^P,m_T^P+1}\big)$, and set $k:=1$
\item While $k < m_T^P + 1$, do
\begin{itemize}
\item Compute
\begin{equation*}
T_P(\tau_k):=t_k\equiv \min_{0\leq i<k<j\leq m_T^P+1}
\left\{\UU_{i,j}-\LLL_{i,j}-\sum_{l=i+1}^{k-1} t_l \right\}
\end{equation*}
\item Let $(i'(k),j'(k))$ be the pair of indices for which 
the minimum above is attained, breaking ties by selecting the
largest $j'(k)$ and corresponding smallest $i'(k)$
\item Set $(i'(l),j'(l)):= (i'(k),j'(k))$ for all $k < l < j'(k)$
\item Set $k := j'(k)$
\end{itemize}
\item Let $T_P$ be $+\infty$ outside $P$ and $V_P := \sum_{k=1}^{m_T^P} t_k$
\item Return $T_P$, $V_P$ and $\{(i'(k),j'(k))\}_{k=1}^{m_T^P}$
\end{itemize}
\end{minipage}
}
\caption{Algorithm \textsc{MaxRev}} \label{figure:maxrev}
\end{center}
\end{figure}

Labb\'e et al. \cite{LMS} give another
polynomial time algorithm for this task but it does
not provide the information we need regarding the active toll-free
subpaths. This information will be
instrumental in generating new paths from $P$ and in obtaining an
approximation guarantee.

\begin{figure}[b]
\begin{center}
\input{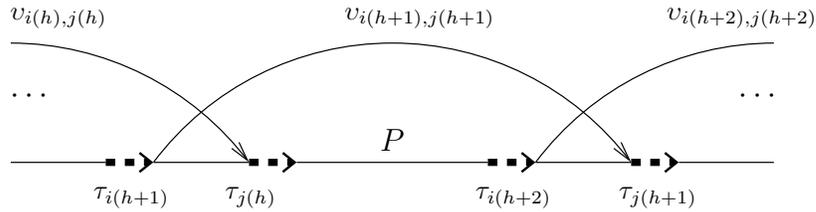}
\end{center}
\caption{A section of a path $P$, showing toll arcs $\tau_{k}$
with $k=i(h+1),\, j(h),\, i(h+2),\, j(h+1)$.}
\label{figure:descendants}
\end{figure}

\begin{theorem}\label{lemma:main}
Let $\upsilon_{i,j}$, $0\leq i<j\leq m_T^P+1$, be the toll-free
subpaths defined in Section~\ref{section:characterizing} and let
$t_k$, $1\leq k \leq m_T^P$ be as in (\ref{eq:optimaltaxes}).
Then, there exists a sequence of paths
\begin{equation}
\upsilon_{i(1),\, j(1)},\, \upsilon_{i(2),\, j(2)},\ldots ,\,
\upsilon_{i(q),\, j(q)}
\end{equation}
with $i(1)=0$ and $j(q)=m_T^P+1$ such that for all $h$ (see
Figure~\ref{figure:descendants})
\begin{equation}
i(h+1)<j(h)\leq i(h+2)< j(h+1)
\end{equation}
and for all $k$
\begin{equation}\label{eq:covering}
|\{h\ :\ i(h)+1 \leq k\leq j(h)-1 \}|\geq 1
\end{equation}
with equality if $t_k\neq 0$.
This implies that for all $h',\, h''$ with $h'\leq h''$
\begin{equation}\label{eq:main}
\sum_{k=i(h')+1}^{j(h'')-1} t_k = \sum_{h=h'}^{h''}  \big[
\UU_{i(h),j(h)} - \LLL_{i(h),j(h)}\big].
\end{equation}
\end{theorem}

\begin{corollary}\label{theorem:profit}
The toll assignment defined by (\ref{eq:optimaltaxes}) is optimal
for $P$.
\end{corollary}

\noindent \emph{Proof of Corollary~\ref{theorem:profit}: } For every
toll assignment $T'$ consistent with $P$
\begin{eqnarray*}
\sum_{k=1}^{m_T^P}T'(\tau_k) &\leq& \sum_{h=1}^q \big[\UU_{i(h),j(h)}-\LLL_{i(h),j(h)}\big],
\qquad \mathrm{by\ (\ref{eq:characterization})\ and\ (\ref{eq:covering})}\\
&=& \sum_{k=1}^{m_T^P}t_k,\qquad \qquad \qquad \qquad \qquad \ \
\mathrm{by\ (\ref{eq:main}).}
\end{eqnarray*}
\qed

\noindent \emph{Proof of Theorem~\ref{lemma:main}:} This proof
is rather cumbersome, so we start by giving an intuitive
description. By Theorem~\ref{theorem:characterization}, 
we know that tolls are constrained by toll-free subpaths.
In order to obtain the corollary above -- which is the ultimate 
purpose of this theorem -- we need
to find a subset of constraints such that each toll arc is covered
by at least one constraint and such that the sum of the slacks of the 
constraints is the same as the sum of the $t_k$'s. Natural candidates
for saturated constraints are derived from (\ref{eq:optimaltaxes}):
when we fix $t_k$, at least one constraint becomes saturated. These
constraints will be denoted by the sequence of indices
$\{(i'(k),j'(k))\}_{k=1}^{m_T^P}$ below. The problem with them is
that they might be redundant. For example, in setting $t_1$ we might
saturate the path $v_{0,3}$ and then in setting $t_3$ we might
saturate the path $v_{0,5}$ which ``supersedes'' $v_{0,3}$. This
situation makes it impossible to obtain equality in
(\ref{eq:main}). To remove redundant constraints, a natural
strategy is to first compute the sequence $\{(i'(k),j'(k))\}_{k=1}^{m_T^P}$,
then to start from $t_1$ and choose among all constraints 
in this subset the one that ``goes furthest'' (say $v_{0,5}$ above). 
Then go to $t_5$ and do the same; and so on. This allows to skip
the redundant constraints. Still, this is not quite what we want 
because there might be nonzero $t_k$'s in the ``overlaps'' of the
chosen constraints (in Figure~\ref{figure:descendants}, 
this would correspond to nonzero $t_k$'s between
$\tau_{i(h+1)}$ and $\tau_{j(h)}$), which again would give strict
inequality in (\ref{eq:main}). It turns out that using the above
strategy but {\em starting from the end} gives the desired properties.
The result of this will be denoted by $\{(i(k),j(k))\}_{k=1}^{q}$
below. To see why this works, consider Figure~\ref{figure:descendants}.
Say that at some point in our choice of saturated constraints
we get to toll arc $\tau_{i(h+2)}$. We then choose the constraint
covering $\tau_{i(h+2)}$ (among $\{(i'(k),j'(k))\}_{k=1}^{m_T^P}$)
which reaches furthest in the direction of $s$, here
denoted $v_{i(h+1),j(h+1)}$. All arcs between
$\tau_{i(h+2)}$ and $\tau_{i(h+1)}$ are now covered (some
of those are not represented in Figure~\ref{figure:descendants}),
so we go to $\tau_{i(h+1)}$. Likewise, we choose the path
$v_{i(h), j(h)}$. Now note that when we set $t_{i(h+1)}$
in the recursion (\ref{eq:optimaltaxes}), the constraint
corresponding to $v_{i(h), j(h)}$ became saturated (or it had
been saturated by a previous toll), so that all toll arcs between
$\tau_{i(h+1)}$ and $\tau_{j(h)}$ (not represented here) 
have been set to 0, as claimed. The heart of the proof that follows is 
to check that this is indeed the case.

\medskip

Consider the following construction. Start with $k=1$.
In the recursive
formula~(\ref{eq:optimaltaxes}), let $(i'(k),j'(k))$ denote
the index for which the minimum is attained when setting $t_k$. 
This makes sense
because we assumed that there exists a toll-free path from $s$ to $t$. In case of
nonuniqueness, select the largest index $j$ and the corresponding smallest
index $i$. Note that there holds
\begin{equation}\label{eq:mainproof2}
t_l=0, \quad \forall\ k+1\leq l \leq j'(k)-1.
\end{equation}
Then rather than evaluating~(\ref{eq:optimaltaxes})  for $t_{k+1}$,
we jump from $t_k$ to $t_{j'(k)}$ and set
\begin{equation}\label{eq:mainproof3}
(i'(l),\, j'(l))=(i'(k),\, j'(k)), \quad \forall\ k+1\leq l \leq j'(k)-1.
\end{equation}
Recursing gives a sequence of indices $\{(i'(k),j'(k))\}_{k=1}^{m_T^P}$.

\medskip

In view of Theorem~\ref{theorem:characterization}, which implies that
\begin{equation}
\sum_{k=i(h)+1}^{j(h)-1}T(\tau_k)\leq \UU_{i(h),j(h)}-\LLL_{i(h),j(h)},
\end{equation}
we say that $\upsilon_{i(h),j(h)}$ \emph{covers} the toll arcs
$\tau_k$  for $i(h)+1 \leq k \leq j(h)-1$.
To derive (\ref{eq:main}), we look for a subset of
$\{\upsilon_{i'(k),j'(k)}\}_{k=1}^{m_T^P}$ that covers all toll arcs of $P$
and such that $t_k=0$ for all arcs $\tau_k$ covered by more than one
 subpath.

\medskip

We proceed backwards. Select $l_1=m_T^P$ and
recursively compute $l_k=i'(l_{k-1})$ until $l_{q+1}=0$ for some
index $q$. There follows:
\begin{equation}\label{eq:mainproof1}
j'(l_1)=m_T^P+1 \quad \mathrm{and} \quad i'(l_q)=0.
\end{equation}
Now, reverse the sequence by setting  $(i(k),\,
j(k))=(i'(l_{q+1-k}),\, j'(l_{q+1-k}))$, $1\leq k\leq q$, to
obtain a sequence that satisfies the assumptions of
Theorem~\ref{theorem:characterization}:

\begin{enumerate}
\item
 $i(1)=0$, $j(q)=m_T^P+1$: This follows from
(\ref{eq:mainproof1}).

\smallskip

\item
 $i(h)<i(h+1)$: This follows from the construction of the
backwards sequence $\{l_i\}_{i=1}^q$.

\smallskip

\item
 $i(h+1)<j(h)$: By contradiction, assume that $j(h)\leq i(h+1)$. This implies that
$\tau_{j(h)}$ is not covered by a toll-free subpath. This is
impossible by the very construction of
$\{(i(k),j(k))\}_{k=1}^{q}$.

\smallskip

\item
 $j(h)\leq i(h+2)$: By contradiction, assume that $j(h)>i(h+2)$.
Then \\$(i'(l_{q+1-h}),\, j'(l_{q+1-h}))=(i(h),\, j(h))$ implies
that $i'(l)=i(h)$ for all indices $l_{q+1-h} < l <
j'(l_{q+1-h})=j(h)$ and in particular for index
$l=l_{q+1-h-1}=i'(l_{q+1-h-2})=i(h+2)$ ($<j(h)$ by assumption).
This implies the contradiction $i'(l_{q+1-h-1})=i(h+1)=i(h)$.

\smallskip

\item
(\ref{eq:covering}) and (\ref{eq:main}): By construction,
every arc is covered by at least one path. By the preceding inequalities,
the only arcs covered by more than one path must belong to the interval
$k=i(h+1)+1,\ldots, j(h)-1$,
for some index $h$. By~(\ref{eq:mainproof2}), their tolls must be
zero, and~(\ref{eq:covering}) is satisfied; (\ref{eq:main}) then
follows from (\ref{eq:covering}).

\end{enumerate}
\qed

\subsection{Partitioning the set of toll arcs}

The approximation algorithm progressively removes toll arcs
from the network. It makes use of the following definition.
\begin{definition}
A {\bf descendant} $P'$ of a valid path $P\in\NN_0$ is a simple
path from $s$ to $t$ in $\NN_0(P)$ that traverses a subset of the toll arcs of
$P$ in the same order as does
$P$.
\end{definition}

Let $P$ be a valid path in $\NN_0$. Theorem~\ref{lemma:main}
suggests a way of constructing a descendant of $P$ that stands a
chance of achieving a high revenue, whenever $P$'s revenue is low.
Let us consider the set
\begin{equation}
\upsilon_{i(1),\, j(1)},\, \upsilon_{i(2),\, j(2)},\ldots ,\,
\upsilon_{i(q),\, j(q)}
\end{equation}
of toll-free paths such that
\begin{equation}\label{eq:inequalities}
0=i(1)<i(2)<j(1)\leq i(3)<j(2)\leq i(4)<j(3)\leq \cdots  \leq
i(q)<j(q-1)<j(q)=m_T^P+1
\end{equation}
and equality~(\ref{eq:main}) holds. If the maximum revenue on $P$
is $B(P)$ then descendants need not be considered, because their upper
bounds are smaller than or equal to $B(P)$. This happens in particular
if $q=1$. We can therefore assume that $q$ is larger than 1.

\medskip

We now consider two descendants: $P_1$ contains all
$\upsilon_{i(h),j(h)}$ with $h$ odd and is composed of arcs of $P$
between them; $P_2$ is constructed in a similar manner,
with even values of the index $h$. For instance, $P_1$
may start in $s$, borrow $\upsilon_{i(1),j(1)}$, take path
$P$ between $\init (\tau_{j(1)})$ and $\term (\tau_{i(3)})$,
 bifurcate on $\upsilon_{i(3),j(3)}$, return to $P$, and so on.
Such a pattern, which is allowed by~(\ref{eq:inequalities}), is
performed by procedure \textsc{TollPartition}
(see Figure~\ref{figure:tollpartition}). Note that $P_1$ and
$P_2$ have no toll arcs in common and that some toll arcs of $P$
belong neither to $P_1$ nor to $P_2$. The rationale behind this
construction is the relationship between the maximum revenue
achievable on $P$ and the upper bounds on the tolls of $P_1$ and
$P_2$ given by Theorem~\ref{lemma:main}.

\medskip

Both these descendants are valid paths. If this were not the case for $P_1$,
there would exist a subpath $\upsilon_{i(h),j(h')}$ ($h<h'$ both
odd) such that $\UU_{i(h),j(h')}$ is strictly smaller than the
length of $P_1$ between $\term (\tau_{i(h)})$ and $\init
(\tau_{j(h')})$ (indeed, a path is valid if and only if
the null toll vector
is consistent with it). Since this length is equal to
$\LLL_{i(h),j(h')}+\TT_{i(h),j(h')}$ by~(\ref{eq:main}) we obtain
that the toll assignment on $P$ is not consistent, a
contradiction. A similar argument applies to $P_2$.

\begin{figure}[h!]
\begin{center}
\framebox{
\begin{minipage}{15cm}
\textbf{Algorithm} \textsc{TollPartition}

\textbf{Input:} a path $P$ and sequence of indices
$\{(i'(k),j'(k))\}_{k=1}^{m_T^P}$

\textbf{Output:} two paths $P_1,P_2$

\begin{itemize}
\item Set $l:=m_T^P$, $r:=1$
\item (Backwards selection of constraints) While $l>0$, do
\begin{itemize}
\item Set $(i''(r), j''(r)):=(i'(l),j'(l))$
\item Set $l:= i'(l)$, $r:=r+1$
\end{itemize}
\item Let $q$ be the last index in the loop above
\item (Reversing the order of the sequence) For all $1\leq k\leq q$, set
\begin{equation*}
(i(k), j(k)):=(i''(q-k),j''(q-k))
\end{equation*}
\item (Constructing the descendants) Let
$P_1$ contain all
$\upsilon_{i(h),j(h)}$ with $h$ odd and be composed of arcs of $P$
between them; let $P_2$ be constructed in a similar manner,
with even values of the index $h$
\item Return $P_1,P_2$
\end{itemize}
\end{minipage}
}
\caption{Algorithm \textsc{TollPartition}} \label{figure:tollpartition}
\end{center}
\end{figure}

\subsection{A detailed example}

\begin{figure}
\begin{center}
\input{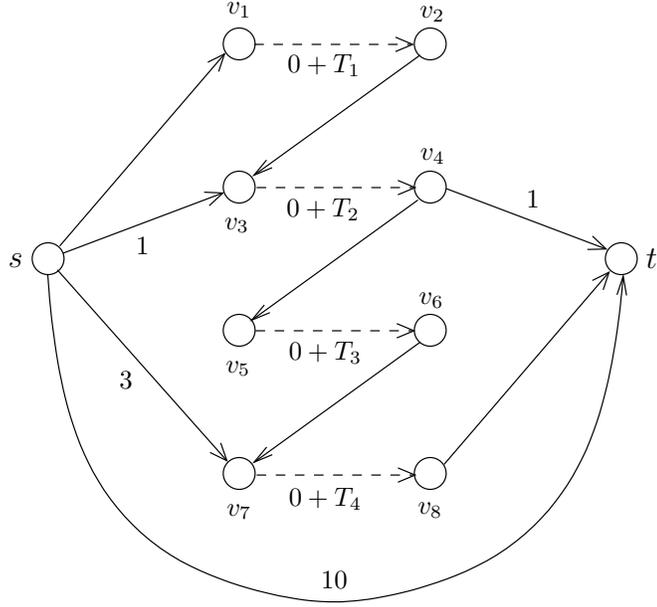}
\end{center}
\caption{An instance of \textsc{MaxToll}. Arcs without label
have cost 0.}
  \label{figure:exemple2}
\end{figure}

We apply the algorithm to the example of Figure~\ref{figure:exemple2}.
Here the shortest path is of length 0 and shortest toll-free path
has length 10 so that $LP=10$.
We start with the shortest path when tolls are set to 0
\begin{equation*}
P_0 = (s,\, v_1,\, v_2,\, v_3,\, v_4,\, v_5,\, v_6,\, v_7,\, v_8,\, t).
\end{equation*}
We apply \textsc{MaxRev}:
$T_1 = T(\tau_1) = 1$ because of $(s,\, v_3)$ (i.e. $v_{0,2}$)
so $(i'(1),j'(1)) = (0,2)$;
$T_2 = T(\tau_2) = 2$ because of $(s,\, v_7)$ (i.e. $v_{0,4}$)
so $(i'(2),j'(2)) = (0,4)$; now we jump over $\tau_3$, setting
$T_3 = T(\tau_3) = 0$ and $(i'(3),j'(3)) = (0,4)$;
finally $T_4 = T(\tau_4) = 1$ because of $(v_4,\, t)$ (i.e. $v_{2,5}$)
so $(i'(4),j'(4)) = (2,5)$. The total profit
is 4. We then run \textsc{TollPartition}. We start
from the end, i.e. $\tau_4$. We get $(i''(1),j''(1)) = (i(4),j(4)) = 
(2,5)$. We jump over to $\tau_2$. We set $(i''(2),j''(2)) = (i(2),j(2)) = 
(0,4)$. And we are done. We finally obtain the sequence 
$(i(1),j(1)) = (0,4)$, $(i(2),j(2)) = (2,5)$ (note in passing
that the only toll arc in the ``overlap'' of 
$(i(1),j(1))$ and $(i(2),j(2))$, i.e. $\tau_3$, has been
set to 0, as claimed by Theorem~\ref{lemma:main}). 
The new paths are thus
$P_1 = (s,\, v_7,\, v_8,\, t)$
and $P_2 = (s,\, v_1,\, v_2,\, v_3,\, v_4,\, t)$. 

\medskip

Now consider
$P_1$. We fix $T_1 = T_2 = T_3 = +\infty$. The only constraint
on $T_4$ comes from the arc $(s,t)$. So we set $T_4 = 10 - 3 = 7$.
There are no descendants to this path.

\medskip

Consider $P_2$. Fix $T_3 = T_4 = +\infty$. Apply \textsc{MaxRev}
again. The toll $T_1$ is bounded by $1$ because of
$(s,v_3)$. Then $T_2 = 10 - 1 - 1 = 8$. The total profit
on $P_2$ is $9$. Again, applying \textsc{TollPartition},
we get that there are no descendants (the descendant is
actually $(s,t)$).

\medskip

Summing up, the best profit was achived on $P_2$ with
$9$. Note
that in this example, the algorithm returns the optimal path, but this
is not always the case.

\section{Analysis of the algorithm} \label{section:analysis}

\subsection{Approximation bound}

We shall prove that \textsc{ExploreDescendants} is an
$\frac{1}{2}\log_2 m_T +1$-approximation algorithm for
\textsc{MaxToll}. The {\sl exact}  approximation factor is given
by the recursion
\begin{equation}\label{eq:factor}
\alpha(k)=\frac{1}{2}\max_{{i+j\leq k}\atop {0<i\leq j < k}}
\{1+\alpha(i)+\alpha(j)\},
\end{equation}
with $\alpha(1)=1$. It can be shown by induction that for all $k$,
\begin{equation*}
\alpha(k) \leq \frac{1}{2}\log_2 k + 1
\end{equation*}

\begin{definition}
The maximum revenue $V_P$ induced by path $P$ is {\bf sufficient}
if
\begin{equation}
V_P\geq \frac{1}{\alpha(m_T^P)}B(P).
\end{equation}
\end{definition}

\begin{theorem}\label{lemma:notsufficient}
Let $P$ be a valid path on $\NN_0$. If the maximum revenue
achievable on $P$ is not sufficient, then either path
 $P_1$ or $P_2$ (say $P'$) returned by \textsc{TollPartition}
satisfies
\begin{equation}
\frac{1}{\alpha(m_T^{P'})}B(P') \geq \frac{1}{\alpha(m_T^P)}B(P)
\end{equation}
\end{theorem}

\noindent\emph{Proof:}
Let
\begin{equation*}
P=\big(\upsilon_{0,1},\, \tau_1,\, \upsilon_{1,2},\, \ldots ,
\tau_{m_T^P},\, \upsilon_{m_T^P,m_T^P+1}\big)
\end{equation*}
with corresponding values of $\UU_{i,j}$, $\LLL_{k,l}$ and
$\TT_{k,l}$ for all $i,\, j,\, k,\, l$, $k<l$. Similarly, for
$r=1,\, 2$,
\begin{equation*}
P_r=\big(\upsilon^r_{0,1},\, \tau^r_1,\, \upsilon^r_{1,2},\,
\ldots ,  \tau^r_{m_T^{P_r}},\,
\upsilon^r_{m_T^{P_r},m_T^{P_r}+1}\big)
\end{equation*}
with corresponding values of $\UU^r_{i,j}$, $\LLL^r_{k,l}$ and
$\TT^r_{k,l}$ for all $i,\, j,\, k,\, l$, $k<l$. Let
\begin{equation*}
\upsilon_{i(1),\, j(1)},\, \upsilon_{i(2),\, j(2)},\ldots ,\,
\upsilon_{i(q),\, j(q)}
\end{equation*}
be the sequence of toll-free paths obtained from
Theorem~\ref{lemma:main}.

\medskip

For all $h>1$, one of the paths $P_r$ contains
$\upsilon_{i(h-1),j(h-1)}$, while the other contains
$\upsilon_{i(h),j(h)}$. Therefore, the subpath of $P$ between
$\term (\tau_{i(h)})$ and $\init (\tau_{j(h-1)})$ is in neither
$P_1$ nor $P_2$. The remaining arcs of $P$ belong to either $P_1$ or $P_2$.
This implies that
\begin{eqnarray*}
\LLL^1_{0,m_T^{P_1}+1} + \LLL^2_{0,m_T^{P_2}+1} &=& \sum_{h=1}^q
\UU_{i(h),j(h)} + \LLL_{0,m_T^P+1} - \sum_{h=2}^q \LLL_{i(h),j(h-1)} \\
&=& \sum_{h=1}^q \big[\UU_{i(h),j(h)} - \LLL_{i(h),j(h)}\big] +
2\LLL_{0,m_T^P+1}
\end{eqnarray*}
By Theorem~\ref{lemma:main}, the first term on the right-hand side is equal
to $\sum_{k=1}^{m_T^P} t_k$ and is strictly smaller than
$B(P)/\alpha(m_T^P)$ by hypothesis. Multiplying by $(-1)$ and
adding $2\UU_{0,m_T^P+1}=\UU_{0,m_T^{P_1}+1} + \UU_{0,m_T^{P_2}+1}$ on both sides, we
get
\begin{equation*}
B(P_1) + B(P_2)>\bigg(2-\frac{1}{\alpha\big(m_T^P\big)}\bigg)B(P).
\end{equation*}
By definition of $\alpha$:
\begin{equation*}
\alpha(m_T^P) \geq
\frac{1}{2}\left[1+\alpha(m_T^{P_1})
+\alpha(m_T^{P_2})\right]  \quad
\Longrightarrow \quad
\bigg(2-\frac{1}{\alpha\big(m_T^P\big)}\bigg) \geq
\frac{\alpha(m_T^{P_1})+\alpha(m_T^{P_2})}{\alpha(m_T^P)}.
\end{equation*}
By substituting in the preceding inequality, we obtain
\begin{equation*}
\alpha(m_T^{P_1})\bigg(\frac{1}{\alpha(m_T^{P_1})}B(P_1)\bigg) +
\alpha(m_T^{P_2})\bigg(\frac{1}{\alpha(m_T^{P_2})}B(P_2)\bigg)
\geq
(\alpha(m_T^{P_1})+\alpha(m_T^{P_2}))\bigg(\frac{1}{\alpha(m_T^P)}B(P)\bigg),
\end{equation*}
which  yields the desired result. Indeed, if we had
$\frac{1}{\alpha(m_T^{P_r})}B(P_r)<\frac{1}{\alpha(m_T^{P})}B(P)$ for $r=1,2$,
this would imply the opposite inequality.
\qed

As a corollary, we obtain the main result of this section.

\begin{corollary}\label{theorem:app}
Let $APP$ denote the revenue obtained from the application of the
procedure \textsc{ExploreDescendants} to a path $P_0$ of shortest
length $\LLL_0$ in $\NN_0$. Then,
\begin{equation}\label{eq:app}
APP\geq \frac{1}{\alpha(m_T^{P_0})}LP\geq \frac{1}{\alpha(m_T)}OPT.
\end{equation}
\end{corollary}
\emph{Proof:} Let $P$ be a valid path and $(\overline{V},\,
\overline{T},\, \overline{P})$ the output of
\textsc{ExploreDescendants}($P$). It is sufficient to show, by
induction on $m_T^P$ that
\begin{equation}\label{eq:property}
\overline{V}\geq \frac{1}{\alpha(m_T^P)}B(P).
\end{equation}
This statement is true if $m_T^P=1$ since the upper bound $B(P)$
is always achievable on a path with a single toll arc. Now assume
that the property holds when the number of toll arcs is less than
$m_T^P > 1$. Let $(V_P,\, T_P):=\mathrm{\textsc{MaxRev}}(P)$. If
$V_P$ is sufficient,  (\ref{eq:property}) is satisfied. If $V_P$
is not sufficient then, by Theorem~\ref{lemma:notsufficient},
\textsc{TollPartition} returns a path $P'$ with
$\frac{1}{\alpha(m_T^{P'})}B(P') \geq
\frac{1}{\alpha(m_T^P)}B(P)$. Since the number of toll arcs in
$P'$ is less than that in $P$, property~(\ref{eq:property}) is
satisfied for $P'$, and the preceding inequality implies that
(\ref{eq:property}) is satisfied for $P$ as well. \qed

Note that
this result applies to the case of negative tolls
as well since the upper
bound~(\ref{UpperBound}) is the same.
Indeed, \textsc{ExploreDescendants} allows only nonnegative tolls
but it computes a feasible solution with a revenue at least
$LP/\alpha$, where $LP$ is unchanged in the unbounded case.

\subsection{Tightness of the approximation} \label{section:gap}

\begin{figure}[t]
\begin{center}
\input{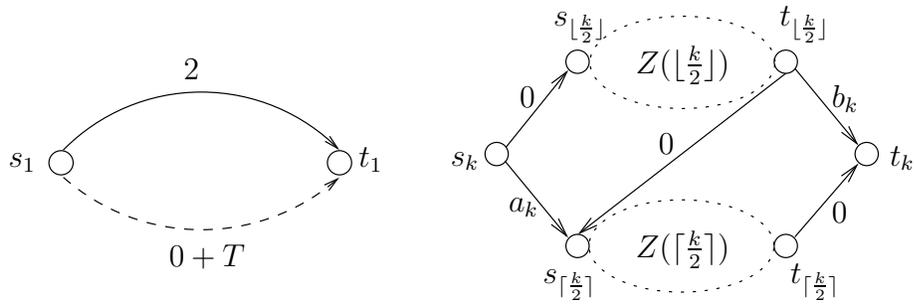}
\end{center}
\caption{Networks $Z(1)$, left, and $Z(k)$, right. Toll arcs are
dashed.}  \label{figure:tight}
\end{figure}

The approximation algorithm determines, in a constructive manner,
an upper bound $\alpha(m_T)$ on the ratio $OPT/LP$. In
this section, we show through a family of instances that this
bound is tight.
\begin{theorem}\label{theorem4}
Let $\mathcal{I}(m_T)$ denote the set of instances of
\textsc{MaxToll}  correponding to a fixed number of toll arcs
$m_T$. Then for all $m_T\geq 1$, the relaxation gap on
$\mathcal{I}(m_T)$ is $\alpha(m_T)$, that is
\begin{equation}
\alpha(m_T) = \max_{I\in
\mathcal{I}(m_T)}\bigg\{\frac{LP[I]}{OPT[I]}\bigg\},
\end{equation}
where $LP[I]$ and $OPT[I]$ denote respectively 
the upper bound and optimal value for instance $I$.
\end{theorem}
\emph{Proof:} Let us consider a two-node
and two-arc network $Z(1)$, with origin $s_1$ and
destination $t_1$. The first arc (a toll arc) has
 fixed cost 0 and the second arc (a toll-free arc)
has fixed cost 2. We recursively construct a network $Z(k)$ made
up of a copy of $Z(\lceil k/2 \rceil )$ and a copy of $Z(\lfloor
k/2 \rfloor )$ (if $k$ is even, there are two copies of $Z(k/2)$),
and two distinguished nodes, the origin $s_k$ and the destination
$t_k$. These are linked by five toll-free arcs, as illustrated in
Figure~\ref{figure:tight}. Strictly speaking, the vertices
$s_{\lceil k/2 \rceil},\, t_{\lceil k/2 \rceil},\, s_{\lfloor k/2
\rfloor},\, t_{\lfloor k/2 \rfloor}$ should be re-labeled,
otherwise many vertices will share the same label even though we
consider them all distinct. The parameters $a_k$ and $b_k$ of
$Z(k)$ are set to
\begin{equation}
a_k=1+\alpha\Big(\bigkfloor\Big) - \alpha\Big(\bigkceil\Big) \quad
b_k=1-\alpha\Big(\bigkfloor\Big) + \alpha\Big(\bigkceil\Big),
\end{equation}
with $a_1=b_1=1$ and $\alpha$ is defined in (\ref{eq:factor}).
Note that $a_k,\, b_k\geq 0$ because $0\leq
\alpha(\lceil k/2 \rceil)-\alpha(\lfloor k/2 \rfloor)\leq 1$.

\medskip

It is not difficult to show, by induction, that
\begin{equation}\label{notdifficult}
\alpha(k)=\frac{1}{2}\bigg[1+\alpha\Big(\bigkceil\Big)+
\alpha\Big(\bigkfloor\Big)\bigg].
\end{equation}
Let $LP(k)$ and $OPT(k)$ denote, respectively, the relaxed and
optimal revenue values on $Z(k)$. Let $\UU(k)$ be the length of
the shortest toll-free path from $s_k$ to $t_k$ in $Z(k)$. We
claim that $\UU(k)=2\alpha(k)$. Indeed, we have
$\UU(1)=2\alpha(1)=2$ and, assuming that $\UU(k')=2\alpha(k')$ for
all $k'<k$, it follows from~(\ref{notdifficult}) that:
\begin{equation}
\UU(k)=\min\bigg\{\UU\Big(\bigkfloor\Big)+b_k,\,
\UU\Big(\bigkceil\Big) + a_k,\,
\UU\Big(\bigkceil\Big)+\UU\Big(\bigkfloor\Big)\bigg\} = 2\alpha(k).
\end{equation}
Clearly, $OPT(1) = LP(1) = 2$ and ${LP(1)}/{OPT(1)}= \alpha(1)=1$.
For $k>1$, the shortest path in $Z(k)$ with tolls set to~0 has
length 0. This implies that $LP(k)=\UU(k)=2\alpha(k)$. To
conclude, we need to show (by induction) that $OPT(k)=2$. We
consider two cases. If the optimal path on $Z(k)$ goes through the
arc joining $t_{\lfloor k/2 \rfloor}$ and $s_{\lceil k/2 \rceil}$,
then $OPT(k)\leq 2$ because $a_k+b_k = 2$. We actually know
in that case that $OPT(k) = 2$ because, by induction,
a profit of 2 is achievable inside each of $Z(\lceil k/2 \rceil )$ and $Z(\lfloor
k/2 \rfloor )$ separately and the only new constraints are those
associated to $a_k$ and $b_k$. 
Otherwise, the optimal
path belongs entirely to $Z(\lfloor k/2 \rfloor)$ or $Z(\lceil k/2
\rceil)$ in which case $OPT(k)=2$ by the induction hypothesis.
\footnote{It is straightforward to check that the same argument
works in the negative tolls case.}
\qed

We have just proved the optimality of our approximation factor
with respect to the upper bound. We can prove more than that: the
same family of instances, under a slight modification, can be used
to show that our analysis of {\sc ExploreDescendants} is tight.
For this purpose, we require instances where $APP$ is much smaller
than $OPT$. This is not the case in the examples of
Figure~\ref{figure:tight} where, indeed, $APP=OPT=2$. However, our
aim is attained if we add to $Z(k)$ a toll arc of fixed cost~1
from $s_k$ to $t_k$. Then $OPT = LP-1 = 2\alpha(k) -1$ (the
optimal path being the new toll arc) yet $APP$ is still 2 since
the algorithm starts with a path of length zero and misses the
optimal path of length 1.

\subsection{Running time analysis}

\textsc{MaxToll} is initialized with a shortest path $P_0$, which can be
computed in $O(n^2)$ time. The toll arcs of the descendants
constitute a subset of the toll arcs in $P_0$, and their traversal
order is the same. Therefore, the values $\UU_{i,j}$, $i<j$
computed for $P_0$ can be reused for all descendants, under an
appropriate renumbering. This operation is achieved in
$O(m_T^{P_0}n^2)=O(m_Tn^2)$ time. The time required to evaluate
$\LLL_{i,j}$, $i<j$, as the algorithm proceeds, is much smaller.

\medskip

Within \textsc{ExploreDescendants}, \textsc{MaxRev} requires at
most $O((m_T^P)^3)$ to compute the maximal revenue induced by path
$P$. Based on the indices $\{(i'(k),j'(k))\}_{k=1}^{m_T^P}$
obtained from \textsc{MaxRev}, \textsc{TollPartition} generates
two descendants in $O(m_T^P)$ time. It follows that the running
time of \textsc{ExploreDescendants} on a path $P$ is determined by
the recursion
\begin{equation}
\mathbb{T}(m_T^P)=\mathbb{T}(m_T^{P_1}) + \mathbb{T}(m_T^{P_2}) +
O((m_T^{P})^3)
\end{equation}
where $m_T^{P_1}+m_T^{P_2}\leq m_T^P$. Therefore, the worst-case
complexity, achieved when $m_T^{P_1}$ is always equal to
$m_T^P-1$, is $O((m_T^P)^4)$. The worst-case running-time of the
entire algorithm is $O(m_T(m_T^3+n^2))$.

\section{Concluding remarks}

Our algorithm can also be applied to the multi-commodity extension
of \textsc{MaxToll} considered in~\cite{LMS}, where each commodity $k\in\cal K$ is
associated with an origin-destination pair. Given a demand matrix,
users solve shortest path problems parameterized by the toll
vector $T$. If distinct tolls $T^k$ could be assigned to distinct
commodities, the multi-commodity extension would reduce to a
$|\cal K|$-fold version of the basic problem. Otherwise,
the interaction between commodity flows on the arcs of
a common transportation network complicates the problem, both from
a theoretical and algorithmical point of view. Of course, we can
obtain an $O(|{\cal K}|\log m_T)$ guarantee by applying \textsc{MaxToll} to
each commodity separately and then selecting, among the $|\cal K|$
commodity toll vectors, the one that generates the highest
revenue. However bad this bound is, we conjecture that it is tight
with respect to the relaxation, which is the sum of the
single-commodity bounds, weighted by their respective demands.
Indeed, we believe that the instances of Figure~\ref{figure:tight}
can be generalized to the multi-commodity case.

\medskip

Other generalizations of \textsc{MaxToll} involve capacity
constraints and lower bounds on tolls.  In the latter case, the
relaxation gap becomes infinite for any value of $m_T$, and our
approach fails, as procedure \textsc{ExploreDescendants} becomes
irrelevant. A completely different line of attack is then
required.

\medskip

Finally, we raise the following important issue: Can our
$\frac{1}{2}\log_2 m_T+1$ guarantee be improved? Such results would
obviously require a tighter upper bound than the one used in this
paper.

\section*{Acknowledgements}

This research was partially supported by NSERC and NATEQ.

\end{document}